# Impact of extended long-range electrostatics on the correlation of liquid-liquid equilibria in aqueous ionic liquid systems

Hugo Marques[*,a], Andrés González de Castilla[b], Simon Müller[b], Irina Smirnova[b]

[a] Centro de Química Estrutural, Institute of Molecular Sciences and Departamento de Engenharia Química, Instituto Superior Técnico, Universidade de Lisboa, Avenida Rovisco Pais, 1049-001 Lisboa, Portugal

[b] Hamburg University of Technology, TUHH, Institute of Thermal Separation Processes, Eißendorfer Straße 38 (O), 21073, Hamburg, Germany

* Corresponding author: hugo.s.marques@tecnico.ulisboa.pt

**Abstract**: Recently an improved long-range model for electrolyte solutions was developed that is applicable from infinite dilution to pure salt. This paper tests this claim for the first time applying it to the calculation of liquid-liquid equilibria for mixtures of different ionic liquids (ILs) and water. The conventional Pitzer-Debye-Hückel (PDH) equation is compared to two of its new, thermodynamically consistent extensions. Both development stages, the extended PDH term and the modified-extended PDH, account for concentration dependent mixture properties instead of using solvent properties. The latter one additionally introduces a modified parameter of closest approach which improves the overall performance of the model for high electrolyte concentrations in systems with variable or low permittivities. To account for the short-range interactions, these long-range models are coupled with the UNIversal QUAsi-Chemical (UNIQUAC) model. Three modeling strategies were tested for the short-range contribution. First, the UNIQUAC parameters were adjusted to each system individually, then the binary interaction parameters were the same for each binary interaction type for all the systems and lastly a linear function of the carbon number was used where possible. For all systems and all modeling strategies tested, the predictive performance increased from PDH to E-PDH and then to ME-PDH. Overall, an introduction of concentration dependent properties and the modification added to ME-PDH enhanced modeling performance when describing these systems, showing the general applicability of this novel long-range term.

**Keywords**: Ionic liquids; Electrolytes; Pitzer-Debye-Hückel; UNIversal QUAsi-Chemical; Binary interaction parameters.

# 1  Introduction

As the interest in salt-based systems grows, the understanding of their thermodynamics becomes of major importance [1–4]. One subset of these compounds are ionic liquids (ILs), which are a class of salts with specific features, such as their high solvation capacity for different types of molecules [5–8], both polar and nonpolar, and their insignificant vapor pressures [5,6,8], which make them potential substitutes to the traditional volatile organic compounds (VOCs) [5,6,9,10]. Generally, ILs are composed of an organic cation and an organic or inorganic anion, usually smaller in size [5,6,10,11]. These are often named "tunable materials" as their properties can be specified for a selected task, simply by choosing the proper combination of ions [5–8,10,11]. Due to the number of ILs possible to be synthesized, characterizing their properties becomes a resource intensive task. As experiments may be time-consuming and costly, one proper tool to evaluate properties of different compounds in mixtures is predictive thermodynamic modeling. When considering IL-based systems, their behavior has been accurately modeled using equations of state (EoS) [12–20], such as the Perturbed Chain based on Statistical Associating Fluid Theory (PC-SAFT) and Cubic Plus Association (CPA), or excess Gibbs free energy ($G^E$) models [21–34], such as the Non-Random Two-Liquid (NRTL) [31,35-38], UNIversal QUAsi-Chemical (UNIQUAC) and UNIversal Function group Activity Coefficient (UNIFAC) [24,26,39–44] or even the predictive Conductor-like Screening Model for Realistic Solvents (COSMO-RS) [29–33,45–48]. Usually, these short-ranged interaction models are coupled with an equation for long-ranged interactions to describe the inter-ionic coulombic correlations present between charged species of ILs. For this type of equations, the most commonly applied long-range terms in chemical engineering are the Debye-Hückel (DH) theory [49], the Mean Spherical Approximation (MSA) [50,51] or the Pitzer-Debye-Hückel (PDH) term [52,53]. Additionally, a Born term is sometimes added to account for the solvation energy of ionic species [14].

Debye-Hückel theory has a hundred years of widespread use and has shown great success in modeling phase equilibria of liquid systems with common salts [53,54], like alkali halides in water. Recently, interest in modeling more complex, concentrated, non-aqueous electrolytes has been growing and many approaches stick to Debye-Hückel theory, showing its applicability [7,17–19,27,31,32,34,44]. However, application in these systems, particularly when highly concentrated, requires consideration of the changing mixture properties, such as density, molar mass and, most importantly [55], the relative permittivity. Based on this, recent works have provided good arguments for the inclusion of a concentration dependent relative permittivity in electrolyte models [18,20,31,32,37,56–65].

The topic of concentration dependent properties has been approached differently by many authors, whether based on the Debye-Hückel perspective [64] or other theories [15,17,56,57], also discussing the role of the Born solvation term [61]. Shilov and Lyashchenko [58,60,63] developed an extended Debye-Hückel theory, while Büllow *et al*. [18] coupled a similar term with the ePC-SAFT EoS. A widely explored alternative is the extended PDH term, occasionally present in literature as ext-PDH [41]. This extension of the conventional PDH equation has been combined with other short-ranged interaction models by several authors. Chang and Lin [31,32] presented its full derivation and coupled it with the COSMO-SAC model tailoring the long-range term to fit ILs in organic solvents. This approach was afterwards tested by Ganguly *et al*. [41] with an eUNIQUAC model. In these cases, concentration dependent properties are employed in ext-PDH and its use outperforms the conventional PDH, especially at high IL concentrations.

More recently, a Gibbs-Duhem consistent modified extension of the PDH term has been published [55,66] following the differentiate down method [67–69]. In the core of its modification, the closest approach parameter $b$ undergoes a semi-empirical scaling. This improves its performance for concentrated aqueous electrolyte systems, but also for electrolytes in organic solvents, for ILs and even for the fused salt state. This is especially observable in very low or strongly variable relative permittivity media. With the inclusion of concentration dependent properties and the aforementioned modification of $b$, it is sought to overcome shortcomings of the conventional PDH. This modified-extended Pitzer-Debye-Hückel (ME-PDH) term was coupled with the predictive COSMO-RS-ES model for the calculation of phase equilibria in several salt systems with pure and mixed solvents, presenting an increase in performance, especially at high salt concentrations and in non-aqueous solvents [55].

Therefore, the scope of this work is to evaluate the performance of all the aforementioned long-range interaction PDH terms (conventional, extended and modified-extended) for mixture of water with several ILs. The short-range interactions are described by the UNIQAC model, which for predictive modeling presents the drawback of having several binary interaction parameters. The model performance is measured by the accuracy of each case in correlating IL solubility in water. One additional objective of this assessment is the study of different trends in IL/water systems in order to reduce the number of adjustable parameters when performing modeling calculations, thereby simplifying the model and reducing computational efforts in the parameterization procedure.

## 2 Theoretical methods

### 2.1 UNIQUAC

The UNIversal QUAsi-Chemical (UNIQUAC) equation is an extension of the quasi-chemical theory of Guggenheim for non-random mixtures to solutions containing molecules of different sizes [70]. It was derived in 1975 by Abrams and Prausnitz and is regarded as a local composition model that considers short-ranged interactions [70,71]. There are two contribution terms: the combinatorial contribution, $G_E^C$, which is an entropic term, as it depends on size and shape asymmetries of the molecules, and the residual contribution, $G_E^R$, which is an enthalpic term, as it is related to the intermolecular forces responsible for the enthalpic effects of mixing. For the UNIQUAC term, the molar excess Gibbs free energy, $G_E^{UNIQUAC}$, is given by [71]:

$$G_E^{UNIQUAC} = G_E^C + G_E^R \qquad (1)$$

Derivation of Eq. (1) yields the activity coefficient for each species $i$ [70] as Eq. (2):

$$\ln(\gamma_i^{UNIQUAC}) = \ln(\gamma_i^C) + \ln(\gamma_i^E) \qquad (2)$$

where $\gamma_i^C$ is the combinatorial term and $\gamma_i^R$ is the residual term of the activity coefficients.

Regarding the combinatorial term, $\gamma_i^C$, it can be calculated using Eq. (3), which is given by the Staverman-Guggenheim expression [70].

$$\ln(\gamma_i^C) = \ln\left(\frac{\phi_i}{x_i}\right) + 5q_i \ln\left(\frac{\theta_i}{\phi_i}\right) + l_i - \frac{\phi_i}{x_i} \sum_{j=1}^{N_{spec}} x_j l_j \qquad (3)$$

where $N_{spec}$ is the number of species in solution, $x_j$ is the molar fraction of component $j$ and the volume fraction, $\phi_i$, and the surface area fraction, $\theta_i$, of component $i$ are determined using Eqs. (4) and (5), respectively.

$$\phi_i = \frac{x_i r_i}{\sum_{j=1}^{N_{spec}} x_j r_j} \qquad (4)$$

$$\theta_i = \frac{x_i q_i}{\sum_{j=1}^{N_{\text{spec}}} x_j q_j} \tag{5}$$

where the parameter $l_i$ can be determined by Eq. (6) [70]:

$$l_i = 5(r_i - q_i) - (r_i - 1) \tag{6}$$

where $r_i$ and $q_i$ are the molecular van der Waals volumes and molecular surface areas, respectively. In Table S1, in the Supplementary Material, values for the structural parameters, $r$ and $q$, for each ion and molecular species can be found. These values are, generally, obtained from the constituents' volume and area group parameters.

The residual term $\gamma_i^R$, given in Eq. (7), is related to the short-range molecular energetic interactions between species in solution and it can be determined using the original expressions of Abrams and Prausnitz [70]:

$$\ln(\gamma_i^R) = q_i \left[ 1 - \ln\left(\sum_{j=1}^{N_{\text{spec}}} \theta_j \tau_{ji}\right) - \sum_{j=1}^{N_{\text{spec}}} \frac{\theta_j \tau_{ij}}{\sum_{k=1}^{N_{\text{spec}}} \theta_k \tau_{kj}} \right] \tag{7}$$

where the parameter $\tau_{ij}$ is given by Eq. (8) [70]:

$$\tau_{ij} = \exp\left(-\frac{\Delta u_{ij}}{T}\right) \tag{8}$$

where $\Delta u_{ij}$ is the UNIQUAC interaction parameter and $T$ is temperature.

The only parameters to be estimated when employing the UNIQUAC model, are binary interactions parameters. These are normally attained from fitting experimental data to the model. Traditionally, each system has an individual set of parameters, even if different systems have some sort of similar interactions. For instance, if two ILs have the same anion interacting with water, it is not common practice to assume that these interactions would be the same as there are more interactions at play. In this work, since the IL is considered as a dissociated cation and anion, rather than a neutral molecule. Therefore, 6 UNIQUAC binary interaction parameters are to be estimated for each system.

## 2.2 Conventional Pitzer-Debye-Hückel

The Debye-Hückel theory [49] is based on the Poisson equation for electrical potential combined with the assumption of a Boltzmann distribution of ion density [49,72]. Due to it not having an analytical solution, it is linearized leading to the Debye-Hückel theory. Within this framework, assumptions like concentration independent relative permittivity and density are made. Additionally in Debye-Hückel theory it is assumed that all ions have the same diameter. The so called PDH term is in essence the Debye-Hückel theory derived via the pressure equation [52]. Both the Debye-Hückel equation and PDH term approach the Debye-Hückel limiting law at infinite dilution [73], but become increasingly inadequate to describe electrolyte systems with rising concentration [72]. Consequently, in the past the purely electrostatic terms were complemented by virial expansions [52,53], where interactions between ions and solvents are described by empirical parameters for the excess Gibbs free energy. The conventional Pitzer-Debye-Hückel (PDH) equation has been successfully applied to aqueous systems composed of common salts and more complex compounds such as ILs [7,17–19,31,32,34,44].

The combination between short and long-ranged interactions models is usually referred to as PDH + UNIQUAC model or eUNIQUAC, which consists of a summation of the excess Gibbs free energies $G^E$ [44,55] as shown in Eq. (9):

$$G_E = G_E^{PDH} + G_E^{UNIQUAC} \tag{9}$$

where $G_E^{PDH}$ is the excess Gibbs free energy contribution from the PDH equation and $G_E^{UNIQUAC}$ is the excess Gibbs free energy contribution from the UNIQUAC model.

The activity coefficients are therefore given by Eq. (10):

$$\ln(\gamma_i) = \ln(\gamma_i^{PDH}) + \ln(\gamma_i^{UNIQUAC}) \tag{10}$$

where $\gamma_i^{PDH}$ is the activity coefficient contribution of the PDH equation and $\gamma_i^{UNIQUAC}$ the activity coefficient contribution of the UNIQUAC model.

The PDH term is given by Eq. (11) [34,53,74].

$$\ln(\gamma_i^{PDH}) = -A_x \left[ \frac{2z_i^2}{b} \ln(1 + b\sqrt{I_x}) + \frac{z_i^2 \sqrt{I_x} - 2I_x^{3/2}}{1 + b\sqrt{I_x}} \right] \tag{11}$$

where $z_i$ is the charge of the species under study and $I_x$ is the ionic strength, in mole fraction basis, given by Eq. (12):

$$I_x = 0.5 \sum_{i=1}^{N_{\text{ions}}} z_i^2 x_i \tag{12}$$

where $N_{\text{ions}}$ is the number of ions in the solution and $z_i$ is the ionic charge of ion $i$.

$A_x$ is the Debye-Hückel parameter, which is calculated using Eq. (13) [53].

$$A_x = \frac{1}{3}\left(\frac{2\pi N_A \rho_0}{M_0}\right)^{0.5}\left(\frac{e^2}{4\pi\varepsilon_0 \varepsilon kT}\right)^{1.5} \tag{13}$$

where $N_A$ is the Avogadro's number, e is the electronic charge, $\varepsilon_0$ is the vacuum permittivity, k is the Boltzmann constant, $\rho_0$ is the solvent's density, $M_0$ is the solvent's molar mass and $\varepsilon$ is the solvent's dielectric constant.

The density and the permittivity of water were described by two polynomials. The function for the density was taken Sato et al. [75] and for the permittivity an own polynomial was fitted to the data from Catenaccio et al. [76]:

$$\varepsilon = 5321T^{-1} + 233.76 - 0.9297T + 0.1417 \times 10^{-2}T^2 \tag{14}$$

$$\rho_0 = 937.029 + 2.566 \times 10^{-3}T + 2.815 \times 10^{-3}T^2 - 7.220 \times 10^{-6}T^3 \tag{15}$$

These equations are valid for all the temperature ranges studied in this work.

Usually, for the closest approach parameter $b$, a constant value is used or it is fitted to the experimental data. It is common practice that this value for monovalent ions is taken as constant $b = 14.9$ [29,37,47,53] which is sometimes applied for a wide range of temperatures, pressures and solvents as a universal parameter, even though it was originally recommended by Pitzer for mixed aqueous metallic nitrates [54]. Additionally, in some sources, the adjusted values $b$ vary considerably [77]. Nevertheless, previous works show that this parameter is related to the distance between centers of the present ions of opposite charges [78,79] and that its considerably large empirical values relate to decaying electrostatics [66], all of which may vary greatly for ILs. Furthermore, $b$ itself does have a physical definition [52] given by Eq. (16):

$$b = a\left(\frac{2e^2 N_A \rho_0}{M_0 \varepsilon_0 \varepsilon kT}\right)^{1/2} \tag{16}$$

where $a$ is an averaged hard-core collision diameter.

As described in previous works [44], the hard-core collision diameter $a$, can be considered as the distance between the centers of present ions. Although this value is salt and solvent specific, averaged values close to 3 Å have shown to be a reasonable average to obtain satisfactory results for many cases [55,66]. For the sake of comparison, this value of 3 Å will also be employed in all PDH based models throughout this work and, in contrast to a previous work [44], is not varied.

## 2.3 Extended Pitzer-Debye-Hückel

Recently, extensions and modifications based on Debye-Hückel theory have been published. One example of this is the work of Shilov and Lyashenko [58,60,63], where a variable relative permittivity is included within the derivation and charging process. An alternative derivation can also be performed for the PDH term.

In this work, an extension of the conventional PDH equation [55] based on the derivation from Chang and Lin [31] will be used. The fundamental differences between this equation and the one derived by Chang and Lin is the application of different mixing rules based on volume fraction for density and permittivity [80,81] and the use of an averaged molar mass for the ions. This extended Pitzer-Debye-Hückel (E-PDH) equation uses the conventional closest approach parameter, while it extends the equation for use of mixture properties instead of solvent properties. The expression for the activity coefficient can be found in Eq. (17) [55].

$$\ln(\gamma_i^{\text{E-PDH}}) = -A_x \left\{ \frac{2z_i^2}{b} \ln(1 + b\sqrt{I_x}) + \frac{z_i^2 \sqrt{I_x} - 2I_x^{3/2}(\frac{M_i}{M_m})}{1 + b\sqrt{I_x}} \right.$$
$$- \frac{2I_x^{3/2}}{(1 + b\sqrt{I_x})} \left[ \left(\frac{V_i}{V_m}\right)\left(1 - \frac{\rho_i}{\rho_m}\right) + \left(\frac{V_i}{V_m}\right)\left(\frac{\varepsilon_i}{\varepsilon_m} - 1\right) \right] \quad (17)$$
$$\left. - \frac{4I_x \ln(1 + b\sqrt{I_x})}{b} \left(\frac{V_i}{V_m}\right)\left(\frac{\varepsilon_i}{\varepsilon_m} - 1\right) \right\}$$

which corresponds to the following mixing rules:

$$\varepsilon_m = \sum_i \phi_i^V \varepsilon_i = \sum_i \frac{x_i V_i \varepsilon_i}{\sum_i x_i V_i} \quad (18)$$

$$\rho_m = \sum_i \phi_i^V \rho_i = \sum_i \frac{x_i V_i \rho_i}{\sum_i x_i V_i} \quad (19)$$

$$M_m = \sum_i x_i M_i \tag{20}$$

$$V_m = \sum_i x_i V_i \tag{21}$$

where $\phi_i^V$ stands for volume fraction of component $i$.

Regarding the electrolyte properties, the molecular mass of each IL can be found in Table S2, in the Supplementary Material. Moreover, a correlation [82] based on the IL's molar mass $M_i$, and molar volume $V_i$, was used to obtain its density as a function of temperature, as described in Eq. (22):

$$\rho_{IL} = \frac{M_i}{V_{m_i}(0.8005 + 6.652 \times 10^{-4}T - 5.919 \times 10^{-4}P)} \tag{22}$$

Where $T$ is the temperature and $P$ is the pressure. In this work, pressure was used as a fixed value of 0.101 Pa, as all experimental data used in this work was obtained at 1 atm. The values of molar volumes used in this work were the ones recommended from the source material of Eq. (22) and can be found in Table S2, in the Supplementary Material.

Since density has a considerably lower impact than the other IL properties [55], such as dielectric constant, the use of a correlation instead of experimental data will not affect deeply the modeled results.

Regarding the dielectric constant, it was calculated using an equation based on experimental data from Singh and Kumar [83] that was fitted to the data of all three different IL families used in this work:

$$\varepsilon_{IL} = -0.2945 n_i + 13.913 \tag{23}$$

where $n_i$ is the length of the alkyl side chain of the methyl-imidazolium ring.

In contrast to the dielectric constant of water, ILs present very low variability in the value of the permittivity in the considered temperature range [55]. Thus, no temperature dependency was applied in this study.

In order to comply with the Debye-Hückel framework [55,81,84], where all ions are exactly equal in every way except for charge, some considerations had to be made for ILs. The molecular mass of each ion is taken as the average of the of the IL's molecular mass:

$$M_i = \frac{M_{IL}}{v} \tag{24}$$

where $v$ is the sum of the stoichiometric coefficients of the ions.

For the other properties of each individual ion, they are considered the same as the IL properties (Eq. (25) and (26)).

$$\rho_{\text{cation}} = \rho_{\text{anion}} = \rho_{IL} \tag{25}$$

$$\varepsilon_{\text{cation}} = \varepsilon_{\text{anion}} = \varepsilon_{IL} \tag{26}$$

## 2.4 Modified-Extended Pitzer-Debye-Hückel

González de Castilla *et al.* [55] developed a modified version of the E-PDH term, where the closest approach parameter undergoes significant changes, i.e. scaling through additional terms added to Eq. (17), that are related to the fluid specification (Bjerrum length divided by $a$). This modified-extended PDH (ME-PDH) is given by Eq. (27) [55]:

$$\ln(\gamma_i^{\text{ME-PDH}}) = -A_x \left\{ \frac{2z_i^2}{b'} \ln(1 + b'\sqrt{I_x}) + \frac{z_i^2 \sqrt{I_x} - 2I_x^{\frac{3}{2}}\left(\frac{M_i}{M_m}\right)}{1 + b'\sqrt{I_x}} - \frac{2I_x^{3/2}}{(1+b'\sqrt{I_x})}\left[\left(\frac{V_i}{V_m}\right)\left(1 - \frac{\rho_i}{\rho_m}\right) + \left(\frac{V_i}{V_m}\right)\left(1 + \frac{b_1}{b'}\right)\left(\frac{\varepsilon_i}{\varepsilon_m} - 1\right)\right] - \frac{2I_x \ln(1+b'\sqrt{I_x})}{b'}\left(2 - \frac{3b_1}{b'}\right)\left(\frac{V_i}{V_m}\right)\left(\frac{\varepsilon_i}{\varepsilon_m} - 1\right) \right\} \tag{27}$$

The modified version of the closest approach parameter $b'$, can be obtained by:

$$b' = b_0 + b_1 \tag{28}$$

In Eq. (28) $b_0$ and $b_1$ are given by Eqs. (29) and (30), respectively.

$$b_0 = \omega_0 b \tag{29}$$

$$b_1 = \left(\omega_1 \frac{\lambda_B}{a}\right)^{3/2} b \tag{30}$$

where $\omega_0 = \frac{3}{2}$ and $\omega_1 = \frac{1}{9}$ are the recommended fixed parameters [55].

The Bjerrum length $\lambda_B$, is defined as the separation at which the electrostatic interaction between two elementary charges is comparable in magnitude to the thermal energy scale. This value is given by Eq. (31):

$$\lambda_B = \frac{e^2}{4\pi\varepsilon_o\varepsilon_m k_B T} \tag{31}$$

In the present work the unsymmetric reference state is applied in all cases. It is noteworthy to mention that symmetric reference state versions of Eqs. (11), (17) and (27) can be found in the literature [55] and could have been applied with equivalent results. Nevertheless, the convention of applying the unsymmetric reference to electrolytes is followed in the present work.

As an overview of the three PDH versions, the different models are tested against each other. Initially the conventional PDH + UNIQUAC is applied to aqueous IL solubility (LLE), where the calculation implies using only the water properties such as molecular mass, density and relative permittivity for long-range electrostatics. Subsequently, LLE calculations are performed again with the E-PDH + UNIQUAC version, where the calculation implies using volume fraction mixing rules and molar mass of the mixture. Lastly, LLE calculations are repeated with the ME-PDH + UNIQUAC version, which also implies consideration of mixture properties in long-range electrostatics and includes an additional term for the calculation of the parameter of closest approach. The gradual increase in complexity is intended to enhance the performance of Pitzer-Debye-Hückel based terms for practical application in highly concentrated IL phases.

## 2.5 Computational Methods

The necessary condition of equilibrium is that, at fixed temperature and pressure, a differential change of composition will not produce any change in the Gibbs energy of the system. However, this is a necessary, but not a sufficient condition of equilibrium between both phases of the liquid-liquid equilibrium, since it does not provide any information about the nature of point, i.e., it may be a maximum, a minimum or an inflection point. Therefore, the usual equilibrium condition is that the chemical potential in both phases must be the same for all components of the mixture. In practice, the criterium used is the isoactivity criterium for each component. Nevertheless, this does not guarantee that the partition coefficient of the cation and the anion are equal. In these conditions, the compositions of each ionic species will be the same, due to the mass balance, but the activity coefficient might not be, which would yield different partitioning coefficient. This encourages the use of mean ionic properties to ensure the partitioning of each ion is the same. For a one-to-one salt dissociation, Eqs. (32) and (33) present the mean ionic mole fraction and the mean activity coefficient for the salt, respectively.

$$x_{Salt} = (x_{Cation} x_{Anion})^{1/2} \tag{32}$$

$$\gamma_{Salt} = (\gamma_{Cation} \gamma_{Anion})^{1/2} \tag{33}$$

Additionally, the partition coefficient $K_i$ depends on the composition of each. Combining it with the isoactivity criterium, Eq. (34) is obtained.

$$K_i = \frac{x_i^I}{x_i^{II}} = \frac{\gamma_i^{II}}{\gamma_i^{II}} \tag{34}$$

where I and II represent the two distinct phases of the LLE. In this work, they are water and ionic liquid-rich phases.

The parameters of each system were optimized using a Differential Evolution algorithm [85]. The strategy used was best bin first, which is a search algorithm design to return the nearest neighbor for a large fraction of queries and a very close neighbor otherwise. Regarding the population size, mutation constant and crossover probability, it was found that the best values that compromise accuracy and computational time are 100, 0.50 and 0.50, respectively. The bounds of search for each interaction parameter were such that the results should provide some physical meaning. It was assumed a limiting value of 50 kJ mol$^{-1}$ in magnitude for the interaction parameters, which means that the values used in this work are bounded between -6000 and 6000 K.

Regarding the objective function employed for optimization, equation (34) was extended to both salt and water, yielding Eq. (35).

$$OF = \frac{1}{n} \sum \left[ \left| K_{Salt}^{exp} - K_{Salt}^{calc} \right| + \left| K_{Water}^{exp} - K_{Water}^{calc} \right| \right] \tag{35}$$

where $n$ is the number of experimental points and the exp and calc superscripts report for the experimental and calculated values, respectively.

To describe the performance of each model, the average relative deviation of compositions ARD was used, which can be calculated using Eq. (36). Taking advantage of this metric, it is possible to display the differences between several models for all the studied systems in a compact way.

$$ARD = \frac{1}{2n} \sum_{i=1}^{n} \left[ \frac{\left| x_{exp}^{IL} - x_{calc}^{IL} \right|}{x_{exp}^{IL}} + \frac{\left| x_{exp}^{W} - x_{calc}^{W} \right|}{x_{exp}^{W}} \right] \cdot 100 \tag{36}$$

## 3 Collected experimental data

Experimental data for 13 systems of IL in water was taken from [44] in order to study the performance of the 3 different models in ILs. The ILs selected are imidazolium-based and the lengths of their alkyl side chain ranges from 2, ethyl, to 10, decyl. These were studied as homologous series, i.e., keeping the anion constant as the side alkyl length varies. However, the effect of the anion was also studied as different anions were selected for this work. These included bis(trifluoromethylsulfonyl)imide, [NTf$_2$]$^-$, hexafluorophosphate, [PF$_6$]$^-$, and tetrafluoroborate, [BF$_4$]$^-$. Table 1 shows the structures of the different homologous series analyzed in this work.

**Table 1: Structures of ionic liquids used in this work. Only the ethyl and decyl-methylimidazolium cations are represented as an indication of the smallest and largest members of the homologous series studied.**

| Ionic liquid family | Anion | Cation |
|---|---|---|
| [C$_n$Mim][NTf$_2$] <br> n = 2-8 | 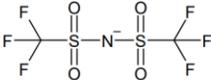 | 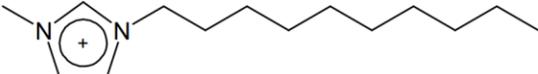 |
| [C$_m$Mim][PF$_6$] <br> m = 4, 6, 8 | 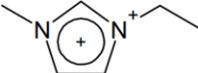 | |
| [C$_k$Mim][BF$_4$] <br> k = 6, 8, 10 | 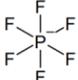 | 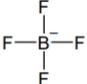 |

The aim of selecting these IL/water systems was also to facilitate the study of different interactions between the components of the mixture. For instance, trends for the anion-water or water-anion interactions along a homologous series may be analyzed. In addition, the effect of different anions can also be observed not only in the binary interaction parameters, but, most importantly, in the modeling results. Furthermore, this feature may be used as a bridge between the traditional methodology of fitting each individual set of parameters for each system and more compact models with less parameters.

Another key point of these aqueous IL systems is that the solubility curves are marginally different from each other when comparing different anions. Moreover, these differences in solubility could show differences between the performance of the models employed in this work as the range of the water and IL-rich phase may be shifted slightly towards different compositions. The liquid-liquid equilibria (LLE) data for all the studied systems in this work can be found in Table S3, in the Supplementary Material.

## 4 Results

As the used objective function does not ensure that, the model actually predicts a LLE, as a first test the Gibbs free energy was plotted as a function of IL mole fraction for one of the systems studied in this work (Figure 1). Using the set of binary interaction parameters obtained from the thermodynamic modeling for this case, it is possible view that the minima of the total Gibbs free energy of Figure 1, represented by the solid line, corresponds to the equilibrium compositions of this system at this temperature. Hence, it can be concluded that both phases of the LLE are correctly calculated with the set of interaction parameters. In some cases, small differences in the experimental and calculated results of the water-rich phase might occur, solely due to extreme low values of IL mole fractions in that phase. Although only the system [$C_6$Mim][$NTf_2$]/water at 313.15 K for the conventional PDH equation is shown, different temperatures, systems and modeling strategies were also tested, yielding the same conclusions as mentioned above. These results are expected, as equations (17) and (27) have been shown to be Gibbs-Duhem consistent in the literature [31,55].

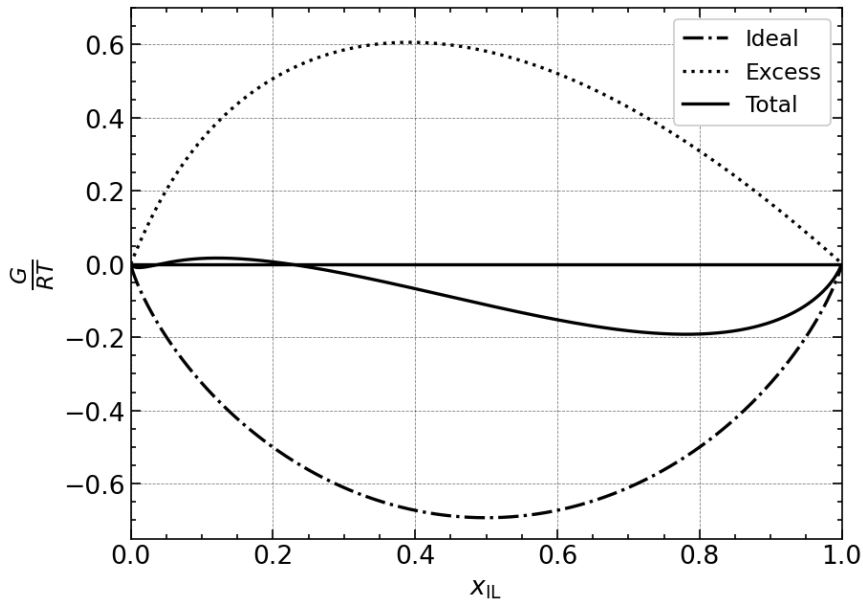

Figure 1: Gibbs free energy plot for [$C_6$Mim][$NTf_2$] in water at 313.15 K. The dot dashed line represents the ideal mixture Gibbs free energy, the dotted line represents the excess Gibbs free energy obtained from the thermodynamic modeling and the solid line represents the sum of both contributions.

### 4.1 Individual Binary Interaction Parameters

The results attained from modeling the abovementioned experimental data for all IL/water systems (cf. Section 3) were compared using an electrolyte UNIQUAC model, where long-range electrostatics was either given by the conventional, extended and modified-extended PDH term.

All calculations for each PDH version can be found in Tables S4 – S6 in the Supplementary Material. In this section, the traditional approach to estimate the UNIQUAC binary interaction parameters was used having one binary interaction parameter for each binary interaction possible in each system independently from all other systems. Therefore, six individual parameters were obtained for each system, since the PDH framework considers the salt as two distinct species, a cation and an anion. Therefore, in order to fit all the 13 studied systems, 78 individual binary interaction parameters are required. These values can be found in Table S7 – S9, in the Supplementary Material. Only the most noteworthy systems are presented in each subsection while the remaining results are presented in the Supplementary Material.

Figure 2 shows the modeling results for [$C_6$Mim][$BF_4$] in water, one of the systems studied in this work where the differences between the different PDH versions are best illustrated. An expected characteristic of the results is that in the water-rich phase (left panel) there are no significant differences between the models. The simple explanation is that the small IL concentration in this phase results in low ion density. Therefore, the three different electrostatic terms tend towards reproducing the Debye-Hückel limiting law, which is the low-concentration limit in all the cases. The finding, that the differences due to the long-range models is mainly observed in the IL-rich phase applies to all studied systems in this work.

Nevertheless, differences become more apparent as IL concentration increases. In the IL-rich phase (Figure 2 – right panel) it can be observed that the UNIQUAC + conventional PDH case attains the lowest accuracy and has difficulty in correlating the curvature of the experimental solubility. In this case, the mixture properties are highly dependent on the IL properties: although molar fraction of the IL is not considerably large, the actual volume fraction of the IL is significant. Thus, the conventional PDH equation performs poorly in non-diluted phases, as it only takes IL-free properties. In contrast, the E-PDH and ME-PDH equations allow a better representation of the phase equilibria. These results support the conclusions of Chang and Lin [31], Ganguly *et al.* [41] and González de Castilla *et al.* [55] regarding superior thermodynamic modeling performance when consistently including concentration dependent properties within long-range electrostatics.

Moreover, although the E-PDH case shows significant improvement over the conventional PDH, the ME-PDH case outperforms the other two versions. Two points can be made here. Firstly, the main feature of E-PDH and ME-PDH is the fact that these seem to provide a more physically meaningful description of phase equilibria at higher electrolyte concentrations. Secondly, the E-PDH equation tends to qualitatively rectify some flaws of the conventional theory, i.e. a better trend when fitting the experimental solubility. However, ME-PDH applies an overall superior correction by damping electrostatic interactions as a function of density, molar mass and (most importantly) relative permittivity. These considerations are valid not only for the system depicted in Figure 2, but also for all the systems studied in this work. Experimental and calculated LLE equilibria for all systems can be found in the Supporting Material.

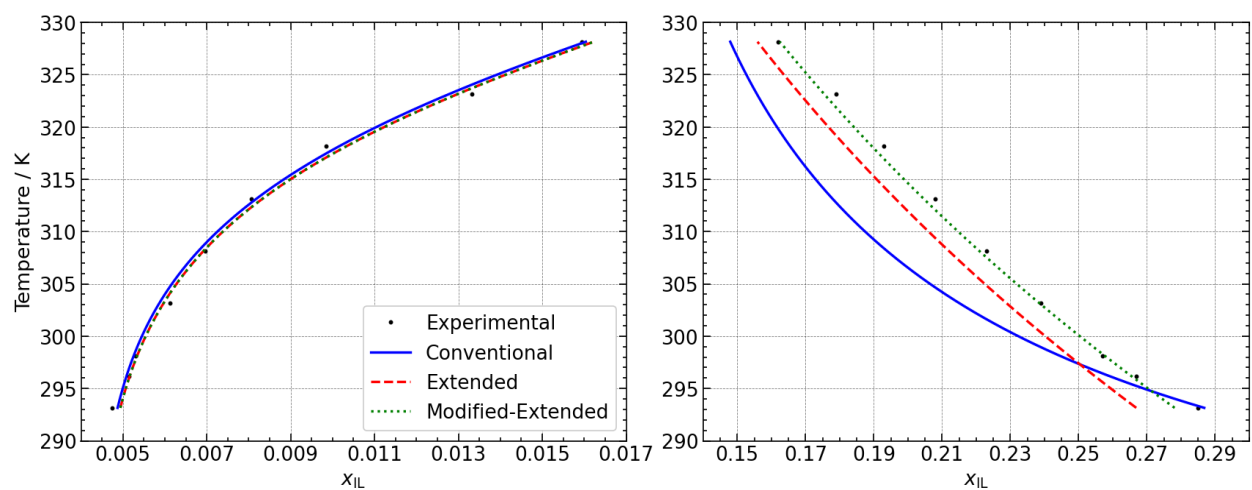

**Figure 2: Solubility curves for [C$_6$Mim][BF$_4$] in water obtained using the PDH, E-PDH and ME-PDH equations coupled with the UNIQUAC model for the water-rich phase (left) and ionic liquid-rich phase (right).**

Similar conclusions were reached regarding the use of the E-PDH and ME-PDH terms coupled with the predictive COSMO-RS-ES model and applied to simpler salts like alkali halides, nitrates, sulfates and perchlorates [55,66]. Thus, the present work additionally also supports the use of the modified parameter of closest approach for ILs.

Another way to observe the difference in performance of models used can be through the composition deviations between the experimental data and modeling results. These are obtained using Eq. (36). The numerical value of the average relative deviation for each system can be found in Table S10, in the Supplementary Material. The errors for the three models for all systems investigated are also summarized in Figure 3.

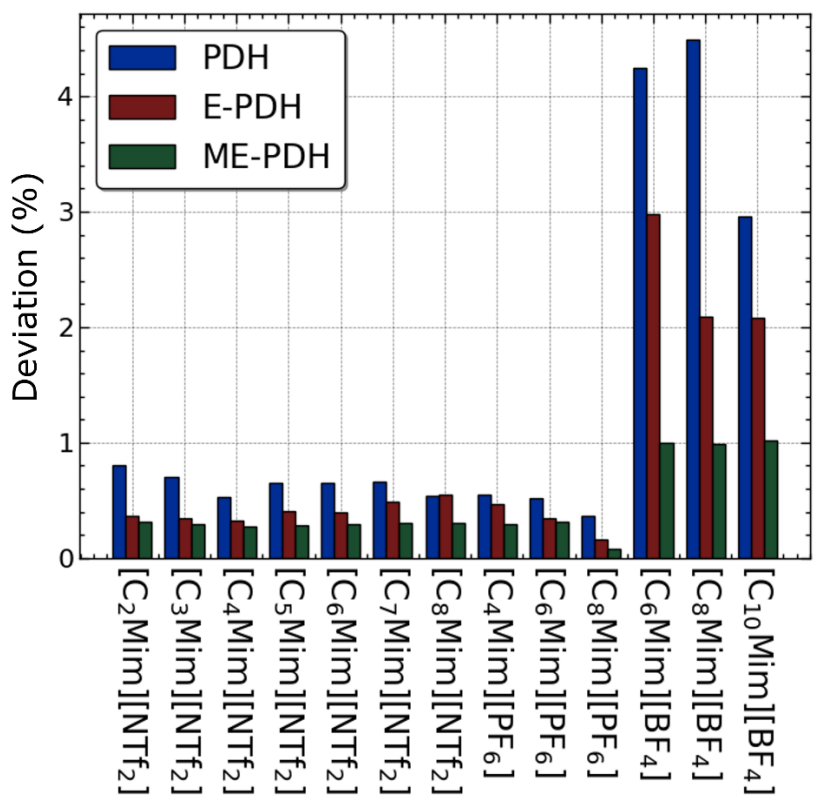

**Figure 3: Composition deviations from experimental data obtained using the PDH, E-PDH, ME-PDH + UNIQUAC model for all the aqueous ILs systems using system specific interaction parameters.**

The improvement of ME-PDH over the other variants can be observed for every IL/water system. For [BF$_4$]-based ILs there is a large improvement of the results while for [NTf$_2$] and [PF$_6$]-based ILs the difference between the modeling approaches is less pronounced. In general, the largest deviations are obtained with the conventional PDH + UNIQUAC model. Furthermore, [BF$_4$]-based ILs show the largest deviations with respect to the other two anion groups. A possible explanation are the anion specific effects on the solubility of ILs in water or even micelle formation as discussed later. For these anions, the miscibility gap is much smaller than for the other two anions (cf. Section 3). The fact that the largest improvements are seen for the systems with the smallest miscibility gap, shows the importance of including a long-range term for the calculation with ILs. Depending on anion type and alkyl tail size of the ions also strong aggregation phenomena in aqueous systems are possible [21,86].

Previous works, such as that of Smith and Robinson [87] verify that for certain models there is an increment of the non-ideal behavior related to increasing non-polar domains because of micelle formation at higher electrolyte concentrations. Chen *et al.* [21] explain this phenomenon as a result of organic ions being larger in size than elemental ions and of the existence of an ionic head group and a tail group present as an alkyl side chain in one of the ions. The coexistence of a hydrophilic component (head group) and a hydrophobic component (tail group) can provide the appropriate conditions for the formation of micelles, which are not accounted for in the present modeling approach.

A comparison to the work of Marques *et al.* [44] is possible as the same ILs in water were studied using a similar model. They only used a temperature dependent density for water but used temperature dependent UNIQUAC parameters. In this manuscript, a temperature dependency was used for the density and also the permittivity, but the UNIQUAC parameters were not temperature dependent. In their work, the performance of a minimum, an average and a maximum value for the closest approach parameter was evaluated. Results from applying the minimum closest approach parameter were selected for comparison, given that this minimum value lies closer to the values obtained for *b* in the present work. When comparing these results from Marques *et al.* [44] with the conventional PDH + UNIQUAC result of the present work, differences lie mainly for systems with the anions $[NTf_2]^-$ and $[BF_4]^-$. The performance is considerably enhanced in the present work for systems with the anions $[NTf_2]^-$ and $[BF_4]^-$ by including the new long-range term. Concerning the $[PF_6]$-based ILs, the smaller members of the homologous series yielded smaller deviations in the work of Marques *et al.* [44] when compared to all models used in this work. For longer alkyl chains, the present approach produced slightly better results.

The trends of the UNIQUAC binary interaction parameters were studied with the aim of simplifying the used terms and minimizing the of number of parameters to be optimized. Two plots of two different trends are presented to clarify the conclusions regarding the evolution of the interaction parameters along the homologous series. The rest of the values can be found in the Supplementary Material.

Figure 4 shows the UNIQUAC binary interaction parameters as a function of the length of the alkyl side chain in [NTf2]-based ILs using the conventional (left panel) and modified-extended (right panel) PDH + UNIQUAC model. For the conventional PDH equation (left panel), trends along the homologous series are distinguishable. However, this is not the case for the E-PDH and ME-PDH equation, the latter shown in the right panel, where trends are not visible. The same conclusion is generally found for the remaining IL families studied in this work ([PF6] and [BF4]-based ILs). Plots for all homologous series in this work using the PDH, E-PDH and ME-PDH equation can be found in Figure S2, in the Supplementary Material.

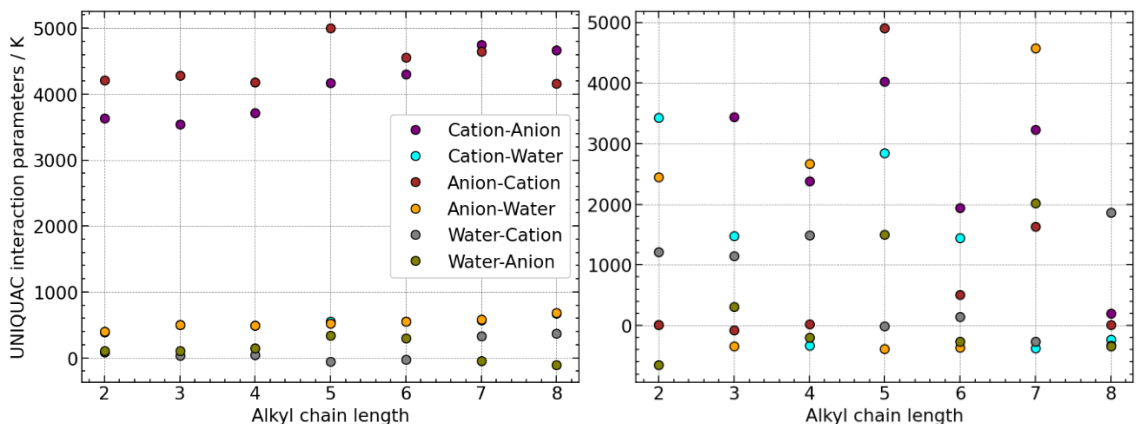

**Figure 4:** Individual UNIQUAC binary interaction parameters as a function of alkyl chain length for [NTf$_2$]-based ILs for the conventional (left) and modified-extended (right) PDH + UNIQUAC model.

As no clear trend can be deduced in many cases for the UNIQUAC system specific parameters, it begs the question if too many parameters are available to fit each system, leading to overfitting of the results. It is possible, that due to the better description of electrostatics, the UNIQUAC contribution is diminished.

### 4.2  Combined binary interaction parameters

The usual approach in thermodynamic modeling with UNIQUAC is to fit systems specific parameters, as done in the previous subsection. In contrast, the present subsection evaluates ion specific parameters to include all experimental data in a single parameterization procedure. By doing so, the number of parameters is reduced: e.g. for [C$_2$Mim][NTf$_2$] to [C$_8$Mim][NTf$_2$] in water, the anion-water and water-anion interactions are assumed to be the same, since water molecules and bis(trifluoromethylsulfonyl)imide anions and water are present along the homologous series. With this change, it is possible to reduce the number of interaction parameters from 78 to 48.

Moving on to the overall accuracy the solubility calculations, conclusions similar to that of the previous approach are reached: the trend in performance of the three models remains the same with PDH + UNIQUAC having the lowest accuracy and ME-PDH + UNIQUAC the highest accuracy. The calculated phase equilibria for all systems using the three models can be found in Table S11 – S13, in the Supplementary Material. Additionally, the values for the 48 combined UNIQUAC binary interaction parameters can also be found in Table S14, in the Supplementary Material.

Even though the number of parameters was reduced, the deviations between the experimental data and the modeling results are not considerably different from the previous approach. The values for the deviation of each system for this modeling approach are shown in Figure 5 and can also be found in Table S15, in the Supplementary Material.

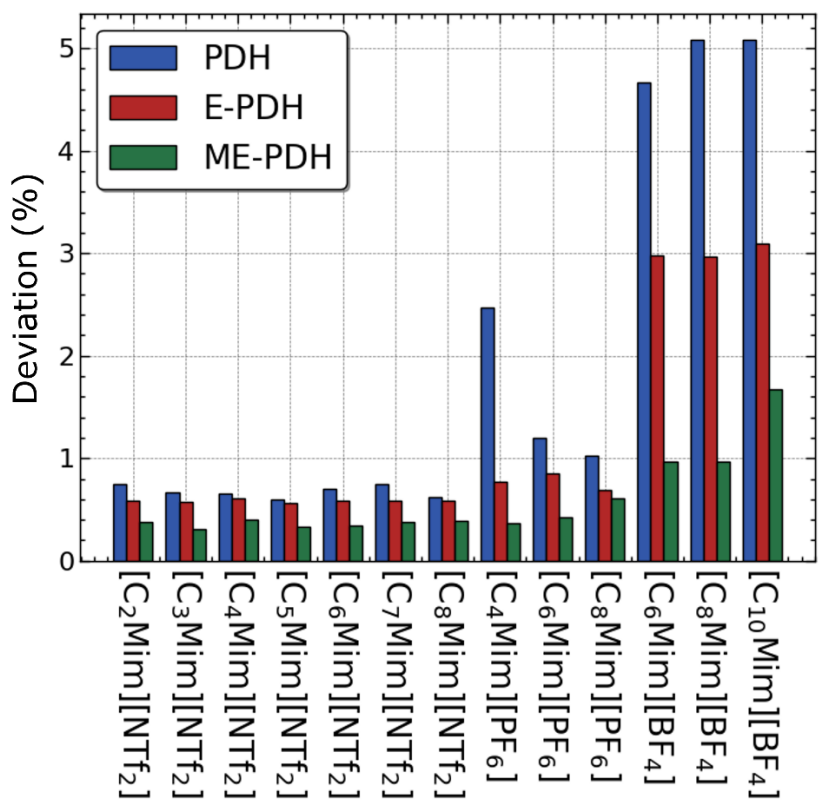

**Figure 5: Composition deviations from experimental data obtained using the PDH, E-PDH, ME-PDH + UNIQUAC model for all the aqueous ILs systems using combined interaction parameters.**

By comparing the results from both approaches, we conclude that the simplification of using combined parameters instead of an individual set of parameters for each IL/water system produces slightly higher deviations. Additionally, the fact that some interactions across different systems were set to be equal may also introduce some degree of error in modeling and a consequent increase in deviations, since in some these systems diverse interactions may have different weights in the performance of short-range interactions. Nevertheless, the general trends obtained previously for the system specific approach were conserved.

Figure 6 shows the interaction parameters for this modeling approach as a function of size of the alkyl side chain for the [NTf$_2$]-based ILs for calculations with the conventional PDH (left panel) and modified-extended (right panel) PDH + UNIQUAC models. In contrast to the individual approach, the behavior of the interaction parameters using the PDH + UNIQUAC model, in the left panel, does not seem to follow any pattern. For the other homologous series using the [PF$_6$]$^-$ and [BF$_4$]$^-$ anions for each model, PDH, E-PDH or ME-PDH, the same conclusions are reached. The plots for all these cases can be found in Figure S3, in the Supplementary Material. The trend of cation-water and water-cation interactions along the homologous series do not seem to have any relation to the alkyl chain length.

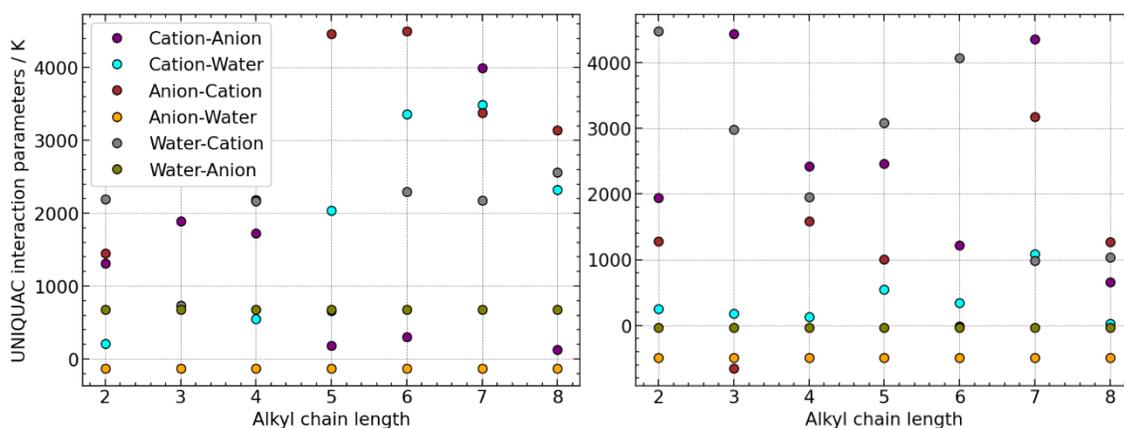

**Figure 6: Combined UNIQUAC binary interaction parameters as a function of alkyl chain length for [NTf$_2$]-based ILs for the conventional (left) and modified-extended (right) PDH + UNIQUAC model.**

Figure 7 presents the UNIQUAC interaction parameters for the anions for each model used in this subsection. An easily identifiable difference in these results is the fact that the conventional water-anion interactions differ considerably from each other and sometimes have positive values that relate to repulsive forces. This is not representative of the reality. The behavior of the attraction of water molecules, mainly by their positive charged hydrogens, to the negatively charged atom of the anion has been studied by several sources, such as Rodrigues *et al.* [88] in the field of Molecular Dynamics (MD). In their work regarding imidazolium cations combined with bis(trifluoromethylsulfonyl)imide anions, it can be seen that water molecules infiltrate the IL and are attracted not only to charged centers, i.e., the imidazolium ring in the cation and the negatively charged nitrogen in the anion, but also to the acidic hydrogens of the cation. Therefore, the resulting parameters remain a mathematical artifact. Figure 7 shows the anions organized by decreasing size, as their volume highly influences their hydrophobic behavior, i.e., their hydrophobic features will be greater as more charge is delocalized. Considering this, it is safe to assume that, in terms of hydrophobic behavior, the trend will be: [NTf$_2$]$^-$ > [PF$_6$]$^-$ > [BF$_4$]$^-$. Having this in mind, only the anion-water interactions for the conventional model and the water-anion interactions for the extended model describe some trends correctly regarding the full physical scope of this approach.

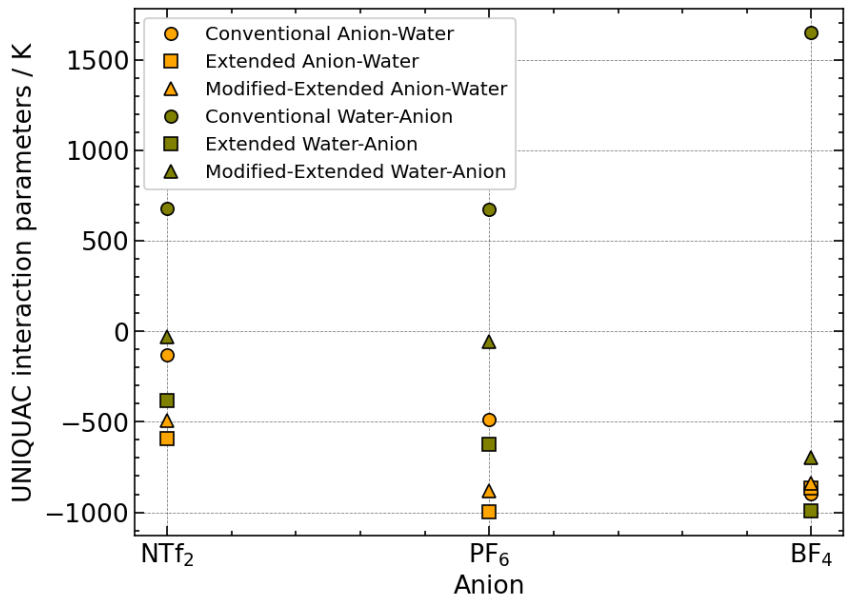

**Figure 7: UNIQUAC binary interaction parameters for the anion-water and water-anion interactions using the conventional, extended and modified-extended PDH + UNIQUAC model for all ions used in this work.**

Ultimately, the restrained grouping of the interaction parameters did not produce considerably higher deviations than the individual parameter approach, while attributing some physical meaning to the interaction parameters.

### 4.3   Linearized binary interaction parameters

The last modeling strategy reduces the amount of needed UNIQUAC parameters even further by describing them as a linear function of the number of carbon atoms in the side alkyl chain. Thus, the cation-water and water-cation interactions along the homologous series were described like so:

$$\Delta u_{C_{n_i} w} = a_{C_{n_i} w} \cdot n_i + b_{C_{n_i} w} \qquad (37)$$

$$\Delta u_{w C_{n_i}} = a_{w C_{n_i}} \cdot n_i + b_{w C_{n_i}} \qquad (38)$$

In Eqs. (37) and (38)  $\Delta u_{C_{n_i} w}$ and $\Delta u_{w C_{n_i}}$ are the cation-water and water-cation interaction parameters for each member of the homologous series with $n_i$ side chain carbon atoms.

A similar approach was also applied for the cation-anion and anion-cation interactions:

$$\Delta u_{C_{n_i}A} = a_{C_{n_i}A} \cdot n_i + b_{C_{n_i}A} \tag{39}$$

$$\Delta u_{AC_{n_i}} = a_{AC_{n_i}} \cdot n_i + b_{AC_{n_i}} \tag{40}$$

In Eqs. (39) and (40) $\Delta u_{C_{n_i}A}$ and $\Delta u_{AC_{n_i}}$ are the cation-anion and anion-cation interaction parameters for each member of the homologous series with $n_i$ side chain carbon atoms. The subscript *A* is used to depict the anion in question, which can either be [NTf$_2$]$^-$, [PF$_6$]$^-$ or [BF$_4$]$^-$.

With these considerations, the initial number of parameters drastically decreases from 78 to 22. All the linearized UNIQUAC binary interaction parameters for each homologous series studied in this work with the PDH, E-PDH and ME-PDH + UNIQUAC models can be found in Table S16 of the Supplementary Material.

By performing additional simplifications to the initial model, concerns regarding the performance of the thermodynamic modeling could be questionable. Nevertheless, as in the previous modeling approaches, the best performing case is the ME-PDH + UNIQUAC model and the one with the largest deviation is PDH + UNIQUAC. The values for the calculated concentration of each aqueous IL system using the three models and this linearized approach can be found in Tables S17 – S19, in the Supplementary Material. In similar manner as the combined approach, the modeling results do not differentiate much from the individual approach.

Figure 8 shows a comparison for the deviation for all IL/water systems using the conventional, extended and modified-extended PDH + UNIQUAC models; their detailed numerical values are presented in Table S20 of the Supplementary Material.

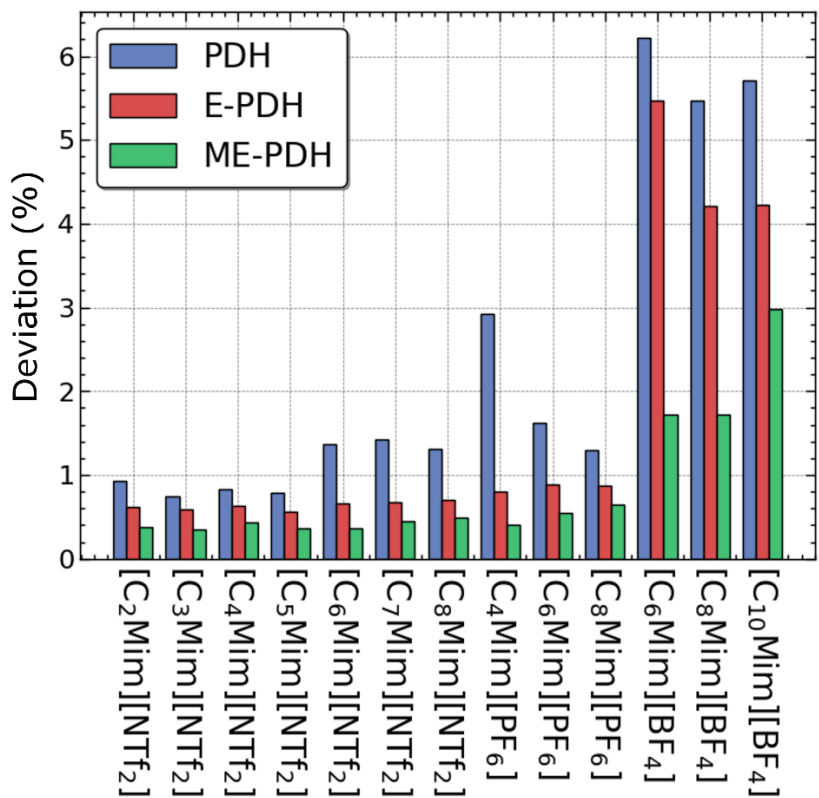

**Figure 8: Composition deviations from experimental data obtained using the PDH, E-PDH, ME-PDH + UNIQUAC model for all the aqueous ILs systems using linearized interaction parameters.**

By examination of the deviations and the modeled solubility, it can be concluded that this approach increased the error mainly in the IL-rich phase, which was also one common feature in the combined approach, yet in this case it is more pronounced. Nevertheless, the deviations are not much higher than the other approaches as they are always under an error of roughly 6% or lower. In order to condense the performance of all the approaches in this work, Figure 9 shows the distinctions between the deviations for all modeling cases with all the aforementioned approaches.

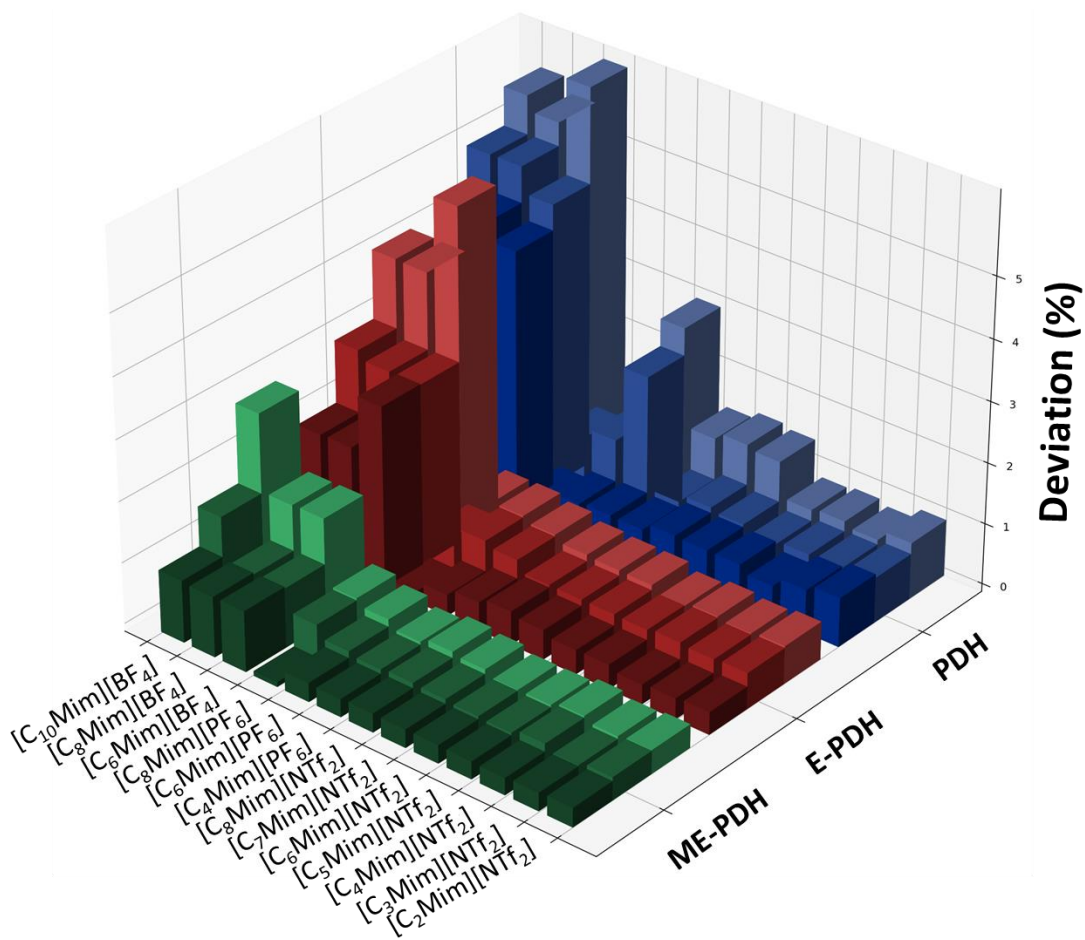

**Figure 9: Differences between the approaches for the UNIQUAC interaction parameters in the modeling of experimental data using the PDH, E-PDH and ME-PDH terms as long-range interaction contribution. The color code used the same as in the previous figures. For each case, the methodology is ordered from left to right as individual - strong color -, combined - lighter color - and linearized - lightest color.**

As expected, there is a moderate increase of the deviations of the models as the number of applied parameters decreases. Furthermore, the number of parameters is not only reduced in a stepwise manner from 78 to 22, but the parameters are also constrained. However, for most cases this does not result in overall strong or outlying deviations. Interestingly enough, the [$BF_4$]-based ILs are where the largest deviations are present. As previously discussed, anion specific interactions may result in water structure-enforced ion pairing [86] and perhaps even micelle formation [21]. Either of these could be the case for [$BF_4$]-based ILs with large alkyl chains. For instance, Figure 10 shows the osmotic coefficient $\phi$ of [$C_4Mim$][$BF_4$] in water at room temperature. Even though [$C_4Mim$][$BF_4$] does not have a considerably large alkyl chain, this [$BF_4$]-based IL evidently does not follow the DHLL at low concentrations, which could be indicative of strong non-idealities that go beyond its sole electrolytic nature. This non-ideal behavior is expected to increase with larger alkyl chains. Indeed, in the vast majority of calculations the deviation also increases with the size of the alkyl chain. It has been shown in the literature that for other $G^E$ models additional considerations are necessary to describe this type of behavior [21].

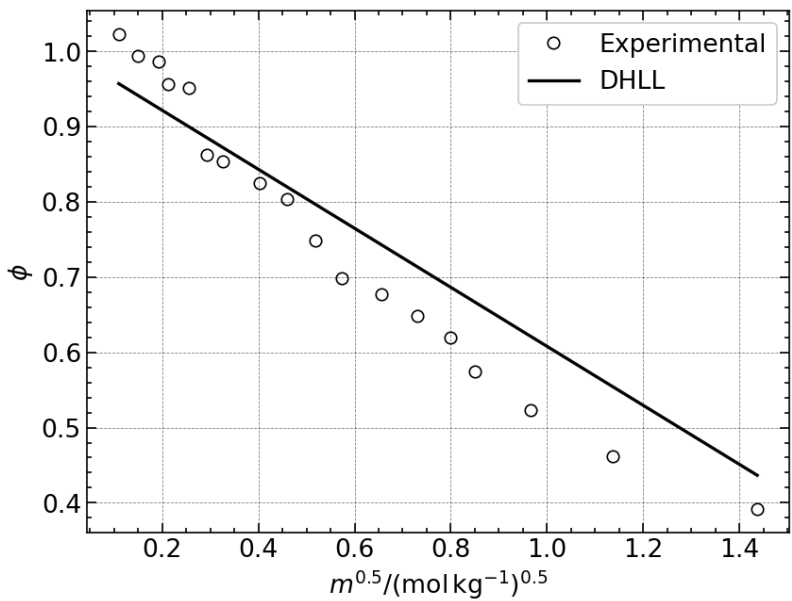

Figure 10. Osmotic coefficient of aqueous [C$_4$Mim][BF$_4$] at 298.15 K as reported by Sheekari and Mousavi [90]. The Debye-Hückel Limiting Law (–) is included as a qualitative referent for electrolyte behavior at low concentrations.

When comparing the different modeling strategies in Figure 9 it can be observed that the E-PDH + UNIQUAC case with linearized parameters (22 in total) performs similarly to the conventional PDH + UNIQUAC case with individual parameters (78 in total). Furthermore, the ME-PDH + UNIQUAC case with linearized parameters outperforms the conventional PDH + UNIQUAC case with individual parameters in most cases. Thus, coupling extended, consistent long-range electrostatics even with less than a third of the original parameters in most cases outperformed the conventional modeling approach.

## 5 Conclusions

The aim of the present work was to analyze conventional and extended versions of the Pitzer-Debye-Hückel equation in aqueous ionic liquid systems. Liquid-liquid equilibria (LLE) of 13 binary systems composed of water and ionic liquids with different anion families were modeled using the UNIversal QUAsi-Chemical (UNIQUAC) model to describe the short-ranged interactions coupled with different Pitzer-Debye-Hückel (PDH) terms for long-ranged interactions. The assessment examines the impact on accuracy resulting from the consistent introduction of mixture properties, such as molecular mass, density and dielectric constant within long-ranged electrostatics.

The resulting models used the well-known conventional PDH equation, which takes salt-free properties; the extended PDH, where the mixture properties were applied with a volume fraction mixing rule; and the modified-extended PDH. In the latter, mixture properties are introduced and the closest approach parameter is scaled as a function of the Bjerrum length to avoid systematic underestimations arising from low and variable relative permittivity values.

No noticeable differences were found between the solubility calculations in the aqueous phase for each model, given the low concentration of ions. Clear differences were observed for the solubility curve deviations between the different models in the IL-rich phase. In the latter phase, the ionic liquid properties become dominant and their consideration improved the accuracy of the calculations. Consequently, the conventional PDH equation produced the least accurate modeling results. The extended PDH equation accounts for concentration dependent properties and introduces a correction. Lastly, the modified-extended PDH, results in the most accurate calculations and outperforms both the conventional and the extended PDH terms. This finding is in agreement with results reported in the literature.

Furthermore, in an attempt to simplify the application of UNIQUAC, the corresponding binary interaction parameters were coupled and reduced. The starting point was the traditional individual fitting of each set of parameters for each system, which results in 78 different interaction parameters. Subsequently, all common interactions between systems are considered, which reduces the number of parameters to 48. Finally in a third approach, the interactions with other species in the mixture are assumed to be linearly dependent of the length of the side chain of the cation. In this linearized approach, the number of parameters is further reduced to 22 thereby constraining the parameterizations. No considerable loss in accuracy was observed and different conclusions were obtained from observation of the trends in the interaction parameters. Remarkably, the performance of the modified-extended PDH equation in the linearized approach with 22 parameters still outperformed even the traditional individual approach when using the conventional PDH equation with 78 system specific parameters.

One concern relates to the formation of non-polar aggregates, water-structure enforced ion pairing and even the potential formation of micelles for bulkier ionic liquids. Even though no large deviations were found that could be clearly attributed to these phenomena, it was identified that tetrafluoroborate based ionic liquids present strong non-ideal behavior even at diluted concentrations and this does impact the overall accuracy in the correlations with respect to the other anion groups.

Concluding, the performance of different models combining the UNIQUAC model with several PDH equations was applied for the successful correlation of the solubility of ionic liquids in water. The modified-extended PDH term was found to provide the most accurate results and it was also found that parameter reduction under reasonable assumptions does not result in significant loss of accuracy for most systems.

## Acknowledgements

Centro de Química Estrutural is a Research Unit funded by Fundação para a Ciência e Tecnologia (FTC) through projects UIDB/00100/2020 and LA/P/0056/2020. This work was funded through project PTDC/QUI-QFI/29527/2017 (including a grant BL36/2022_IST-ID to H. Marques).